# 基于数据科学方法的岩石行星地幔储水能力的研究：地球、火星和系外行星

董俊杰[1,2]

1. 哈佛大学 地球与行星科学系，马萨诸塞州 坎布里奇 02138；2. 哈佛大学 科学史系，马萨诸塞州 坎布里奇 02138

**摘　要**：名义无水矿物(NAMs)是岩石行星地幔中水的主要载体，研究地幔中主要 NAMs 的水溶解度可以帮助我们估测岩石行星地幔理论上的储水能力，并间接约束其内部的实际含水量。通过统计和统计学习算法等数据科学方法，本文旨在介绍并总结前人对地球、火星和系外行星地幔储水能力的模拟研究工作。本文首先回顾了岩石行星地幔储水能力的热力学理论模型；然后围绕地球和火星的两个实例研究，探讨了如何使用包括稳健回归拟合、蒙特卡罗方法和自助聚集算法等方法，把原子尺度的水溶解度实验数据点和其测量误差转化成行星尺度的储水能力模型；接着介绍了系外行星观测的大样本数据是如何帮助我们理解系外岩石行星储水能力的统计学性质；最后讨论了数据科学方法在矿物物理学研究中的局限性，并对如何更好地将统计及统计学习算法与矿物物理数据研究相结合作出展望。

**关　键　词**：名义无水矿物(NAMs)；深部水循环；矿物物理学；热力学；行星科学；数据科学

## A data science approach to study the water storage capacity in rocky planet mantles: Earth, Mars, and exoplanets

DONG Jun-jie [1,2]

1. *Department of Earth and Planetary Sciences, Harvard University, Cambridge, Massachusetts 02138, USA;*
2. *Department of the History of Science, Harvard University, Cambridge, Massachusetts 02138, USA*

**Abstract**: Nominally anhydrous minerals (NAMs) are the primary carriers of water in rocky planet mantles. Therefore, studying water solubilities of major NAMs in the mantle can help us estimate the water storage capacities of rocky planet mantles and indirectly constrain the actual water contents of their interiors. By using the data science methods such as statistics and statistical learning algorithms, in this paper, current modeling studies on the mantle water storage capacities of Earth, Mars, and exoplanets have been especially introduced and summarized. Firstly, the thermodynamic model for mantle water storage capacity has been reviewed. Then, based on the two case studies on Earth and Mars, how to translate atomic-scale experimental data of water solubility and their measurement errors into planetary-scale models of mantle water storage capacity has been explored by using the robust regression, Monte Carlo methods, and bootstrap aggregation algorithms. Thirdly, how the large sample data from the exoplanet observational campaigns can help us understand the statistical properties of their mantle water storage capacities of rocky exoplanets has been introduced. Finally, the application limitations of data science methods in mineral physics research have been discussed, and how to better combine statistics and statistical algorithms with mineral physics data research has been prospected.

## 0 引言

岩石行星的水对其内部演化和表面宜居性有着深远影响。在岩石行星新生吸积的过程中,水被储存到其深部(Morbidelli et al., 2000; Brasser, 2013; Raymond et al., 2022)。随后,岩浆去气作用把包括水及其他大量挥发物质被带到行星表面并形成了早期的海洋与大气(Elkins-Tanton, 2011)。当地球上的板块运动开始后,俯冲板片可能会把表面海洋的水重新带回到其深部(Van Keken et al., 2011),从而形成有效的深部水循环。然而,火星缺乏板块俯冲作用,持续的岩浆去气作用导致其地幔水的净流出,从而使火星内部的实际含水量较低(Dong et al., 2022)。这些岩石行星内部水的实际含量,不仅控制着地幔的流变学性质和温度演化(Nakagawa and Iwamori, 2017),而且通过影响地幔熔点调节火山的排气(Katz et al., 2003),从而决定水在行星表面和内部的分配(Hirschmann, 2006; Karato et al., 2020)。虽然目前对于行星内部在不同演化阶段的实际含水量还无法准确测量,但是,我们可以通过计算其地幔理论上的储水能力,从而间接地研究岩石行星内部含水量。

岩石行星地幔中的名义无水矿物(NAMs),例如橄榄石,其晶体结构中,通常有微量的、以各种形式的氢缺陷存在的结构水(杨晓志和李岩,2016)。这些NAMs都有一个热力学限制的水溶解度,在晶体中的OH饱和时,额外的水就会以自由水、含水熔体及含水矿物(hydrous minerals)的形式存在。因此,一个岩石星球地幔的储水能力由它的矿物相组成和这些NAMs各自的水溶解度决定(Hirschmann et al., 2009)。近几十年,研究人员通过高温高压实验对地幔主要矿物包括上地幔的橄榄石、过渡带的瓦兹利石和林伍德石、以及下地幔布里奇曼石的水溶解度进行了测定[见Dong等(2021)的汇编数据集]。然而,这些在特定高温高压下采集的水溶解度数据点往往相对离散,并且误差较大、自洽性低,导致我们对温度、压强和化学成份这些因素对水溶解度的影响程度仍然不是很清楚,再加上相关的传统拟合公式也不够完善(Hirschmann et al., 2005; Keppler and Bolfan-Casanova, 2006),因此仅仅依靠传统方法我们无法准确地计算像地球和火星这类岩石行星地幔的理论储水能力。然而,统计和统计学习等一系列数据科学分析方法为岩石行星地幔的储水能力的准确计算带来了新视角(Dong et al., 2021, 2022; Guimond et al., 2023)。

本文将首先回顾岩石行星地幔储水能力的热力学理论模型,然后,笔者通过自己围绕地球和火星的两个实例研究(Dong et al., 2021, 2022),探讨如何将包括稳健回归拟合、蒙特卡罗方法和自助聚集算法在内的数据科学方法应用在实验数据的分析和行星储水模型的建立上,并且通过Guimond等(2023)对笔者模型的应用,介绍利用系外行星观测的大样本数据来理解岩石行星储水能力的统计学性质的方法。最后,讨论数据科学方法在矿物物理学研究中的局限性,并对如何将统计和统计学习方法与矿物物理数据结合作出展望。

## 1 岩石行星地幔储水能力的热力学理论

### 1.1 名义上无水矿物(NAMs)的水溶解度

微量的水通常以各种形式的氢缺陷储存在NAMs中。从热力学的角度看,NAMs获取水的过程可以被理解为一种特殊的水合反应,即水($H_2O$)中的质子和缺陷周围尚未质子化的氧原子(O)结合形成OH,从而保持矿物晶体内部整体的电荷守恒(Keppler and Bolfan-Casanova, 2006):

$$\frac{n}{2}H_2O + \frac{n}{2}O^{2-} = (OH^-)_n \tag{1}$$

这种水合反应的平衡常数$K$,可以用方程(2)表示:

$$K = \frac{a_{(OH^-)_n}}{f_{H_2O}^{n/2} \cdot a_{O^{2-}}^{n/2}} \tag{2}$$

式中,$a_{O^{2-}}$和$a_{(OH^-)_n}$分别为未质子化的O和OH的活度,$f_{H_2O}$是$H_2O$的逸度,$n$是水逸度指数,其中$f_{H_2O}$可以用水的状态方程计算(Otsuka and Karato, 2011)。

当近似假设:①晶体结构里未质子化的氧原子的活度约为1($a_{O^{2-}} \sim 1$);②OH只是作为一种浓度极低的溶质存在于NAMs中,其活度可以近似为其浓度[$a_{(OH^-)_n} \sim c_{(OH^-)_n} \sim n \cdot c_{OH^-}$],那么在$H_2O$饱和条件下,NAMs中的OH浓度上限便可以表示为:

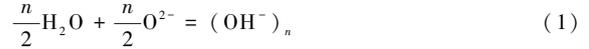

$$c_{OH^-} = \frac{1}{n} f_{H_2O}^{n/2} K \tag{3}$$

同时,处于等温等压平衡态下的标准自由能改变量,$\Delta G^*(P,T)$,与平衡常数$K$相关,根据范特霍夫方程(Van't Hoff equation)可得方程(4):

$$\Delta G^*(P,T) = -RT\ln K = \Delta H^* - T\Delta S^* + \Delta V^*(P - P_0) \tag{4}$$

式中,$\Delta H^*$和$\Delta S^*$分别为参考压强($P_0$)下的反应焓和反应熵,$\Delta V^*$是水合反应下反应物和生成物的

体积差,$R$ 是理想气体常数。因为大多数地幔矿物富集镁,所以这里选用各矿物的镁端元(Mg end-member)作为方程(4)中的标准态 $\Delta G^*$。

当然,这些矿物中的其他成分也会影响吉布斯自由能变化。例如,水合反应的吉布斯自由能变化 $[\Delta G(P,T,X_{Fe})]$ 实际上会随着铁含量($X_{Fe}$,以摩尔分数为单位)的改变而改变。如果我们把这些矿物看作是镁和铁的理想固溶体,通过麦克劳林展开式,我们可以将这个反应的实际吉布斯自由能变化则为方程(5)(Schmalzried,1995;Zhao et al.,2004):

$$\Delta G(P,T,X_{Fe}) \sim \Delta G^*(P,T) + X_{Fe}\Delta G^*(P,T)' + 高阶项 \quad (5)$$

式中,$X_{Fe}\Delta G^*(P,T)'$ 表示铁的存在所产生的实际吉布斯自由能变化与镁端元标准态的偏差。由于铁含量相对较低,所以方程(5)里的高阶项可以忽略不计。通过结合方程(3)至方程(5),NAMs 中的 OH 浓度上限($c_{OH}$,以摩尔为单位)可表示为:

$$c_{OH} = \frac{1}{n}\exp\left(\frac{\Delta S^*}{R}\right) \cdot \exp\left[-\frac{\Delta H^*}{RT} - \frac{\Delta V^*(P-P_0)}{RT}\right] \cdot$$
$$\exp\left[-\Delta G^*(P,T)'\Delta\frac{X_{Fe}}{RT}\right] \cdot f_{H_2O}^{n/2} \quad (6)$$

$$c_{H_2O} = \frac{1}{2}c_{OH} \quad (7)$$

在 $H_2O$ 饱和条件下,进一步简化方程(6)和(7),NAMs 中的水溶解度(Dong et al.,2021,单位为质量的 % 或 $\times 10^{-6}$)则可通过方程(8)计算得到:

$$\ln(c_{H_2O}) = \ln\left(\frac{1}{2n}\right) + a + \frac{n}{2} \cdot \ln(f_{H_2O}) + \frac{b + c \cdot P + d \cdot X_{Fe}}{T} \quad (8)$$

式中,$a$、$b$ 和 $c$ 分别是与反应的熵、焓和体积变化有关的常数,$d$ 与成分效应有关。$T$ 为温度(K),$P$ 为压强(GPa)。在与地幔有关的条件下,$P \gg P_0$,$P_0$(通常为一个大气压)变得可以忽略不计,即,$P-P_0 \sim P$。

严格来讲,NAMs 的水溶解度同时受多种内在和外在因素控制:前者主要是矿物的缺陷类型与丰度,受矿物结构、元素组成、水的耦合机制等因素控制;后者主要是环境参数,包括温度、压强、氧逸度、共存矿物组合、流体相等因素(杨晓志和李岩,2016)。本节中介绍的热力学模型主要考虑了温度、压强和铁含量对 NAMs 水溶解度的影响。在对目前已有的矿物物理实验数据进行拟合后,该模型可以较为简练地定量描述水溶解度和它们的半经验关系。但是,这类半经验模型仍有以下不足之处:①氧逸度($f_{O_2}$)尚未被包括在该热力学模型中。$f_{O_2}$ 对地幔主要 NAMs 的水溶解度可能有较大影响。比如说,$f_{O_2}$ 可以通过改变 Fe 的价态,影响矿物中 OH 耦合的电荷平衡,进而影响瓦兹利石、林伍德石等矿物的水溶解度(Withers and Hirschmann,2008;Gaetani et al.,2014;Yang,2016;Blanchard et al.,2017;Fei and Katsura,2020;Liu and Yang,2020;Druzhbin et al.,2021;Zhang et al.,2022)。②微量元素效应尚未被包括在该模型中。比如说,Ti 影响橄榄石的水溶解度(Padrón-Navarta and Hermann,2017),Al 影响布里奇曼石的水溶解度(Ishii et al.,2022),以及 F 可能影响石榴石水溶解度(Crépisson et al.,2014)。③在 NAMs 的晶体结构中,尽管氢可以以包括羟基(OH)、水分子($H_2O$)、分子氢($H_2$)、氨根离子($NH_4^+$)等多种形式存在(Hirschmann et al.,2012;Yang et al.,2016),但该模型将它们笼统地概括成 OH。

### 1.2 地幔矿物平衡相组成及其储水能力

地幔矿物相组成和这些矿物各自的水溶解度决定了一个岩石星球地幔的储水能力。在过往的研究中,对地幔整体储水能力的估算并没有全面考虑地幔矿物相组成和其水溶解度随着温度和压强的变化(Hirschmann et al.,2005)。通过使用基于最小吉布斯自由能原理的热力学软件(如 HeFES-To,Stixrude and Lithgow-Bertelloni,2011;PerpleX,Connolly,1990),Dong 等(2021,2022)和 Guimond 等(2023)计算了岩石行星内部地幔矿物平衡相组合沿着地热梯度的复杂变化(不包括熔化),并在对这些矿物相组合储水能力的计算中考虑了温度和压强的变化所产生的影响。二十多种常见的地幔矿物相被考虑在其中,包括橄榄石、瓦兹利石、林伍德石、布里奇曼石、辉石、石榴子石、毛钙硅石和铁方镁石等。

在计算岩石行星地幔的整体储水能力时,我们首先根据实验所得的温度、压强和铁含量数据对橄榄石、瓦兹利石和林伍德石的水溶解度进行拟合(在第二、三节具体讨论)。然后,根据这三种主要矿物(i)相对于次要的地幔矿物相(j)的水分配系数($D_{H_2O}^{i/j}$),计算包括辉石和石榴子石这些次要矿物相的水溶解度及其随温度压强的变化:

$$c_{H_2O}^j = c_{H_2O}^i / D_{H_2O}^{i/j}(T,P,X_{Fe},X_{Al}) \quad (9)$$

式中,$D_{H_2O}^{i/j}$ 是 $T$、$P$、$X_{Fe}$ 等的函数。接着,可以对在特定温压下单个矿物相的水溶解度进行计算:

$$c_{H_2O}^i = c_{H_2O}^j / D_{H_2O}^{i/j}(T,P,X_{Fe} 等) \quad (10)$$

结合特定温压下各个矿物平衡相的相对丰度($X_{i,j}$),该温压条件下地幔地储水能力可通过各矿物水溶解度的加权和得到:

$$c_{H_2O} = (c_{H_2O})_i \cdot [X_i + \sum_j (X_j \cdot D_{H_2O}^{i/j})] \tag{11}$$

而在特定地幔潜在温度下地幔的整体储水能力,最终可以通过沿着其地温梯度的地幔储水能力曲线进行积分而计算得到(Dong et al., 2021, 2022)。

需要明确的是,对于真实体系(自然界)的地幔矿物相组合来说,其实际储水能力往往低于方程11所描述的所有矿物相水溶解度的简单加权和(Ardia et al., 2012; Férot and Bolfan-Casanova, 2012; Tenner et al., 2012)。严格来讲,笔者所指的水溶解度只能被看作是在理想条件下地幔储水能力的理论上限。另外,NAMs 水溶解度受共存流体相影响。该模型假设氢为流体相中的唯一挥发分,然而真实的地幔流体是 C、H、O 等的复杂混合物。我们对复杂流体相成分对 NAMs 水溶解度以及地幔储水能力影响的理解还十分有限。

## 2 基于稳健回归拟合和蒙特卡罗方法的地球地幔储水能力模型

近几十年来,学界对 NAMs 水溶解度的高压实验研究实现了大量成果突破。基于本文的主旨是探究数据科学方法的应用,笔者略去对这些实验成果的全面回顾,具体的实验测量方法和过程可以参照杨晓志和李岩(2016)。笔者只在这里简单讨论对 NAMs 水溶解度现存数据集的整理和修正。

由于高压实验样品小、含水量标定困难的问题,现有的 NAMs 水溶解度数据点往往相对离散,并且误差较大、自洽性低。例如,被广泛使用的含水量标定结果(Paterson, 1982),是根据硅酸盐玻璃和石英等多种材料建立的粗略模型。当我们用它来测量含水量时,误差有时可以超过150%(Bell et al., 2003)。再比如,目前用傅里叶变换红外光谱(FTIR)测得的含水量数据大多是使用非偏振光(unpolarized light)对光学各向异性的多晶矿物样品进行测量的结果,但这些结果自洽性低(Libowitzky and Rossman, 1997)。近些年,Bell 等(2003)和Withers 等(2012)对橄榄石中的含水量进行了相对准确的红外吸收定标,基于 Bell 等(2003)的交叉标定,Dong 等(2021)先将基于 Paterson(1982)或 Libowitzky 和 Rossman(1997)标定测得的含水量向上修正了2~4倍,然后再根据 Withers 等(2012)的交叉标定将基于 Bell 等(2003)的实际值修正为其原始值的2/3。对于瓦兹利石和林伍德石,Dong 等(2021)根据 Bolfan-Casanova 等(2018)基于 ERDA(elastic recoil detection analysis)的定性研究,将基于 Paterson(1982)测得的瓦兹利石和林伍德石水溶解度的实际值修正为原始值的181%(Bolfan-Casanova et al., 2018)。修正这些数据的目的是减少现有数据集中已知的系统误差(systematic error),以确保汇编的水溶解度的数据具有一定内部自洽性。测量含水量的另一种方法——SIMS(secondary ion mass spectrometry),其关于橄榄石的数据有较好的自洽性(Bolfan-Casanova et al., 2018),所以 Dong 等(2021)未对汇编的 SIMS 数据进行校正。需要注意的是,无论 ERDA 还是 SIMS,测定的都是样品中各种形式氢(OH、$H_2$、流体包裹体甚至有机氢)的总含量,并不一定是 OH 含量,这与 FTIR 有所区别。

对现有数据集中已知的系统误差进行尽可能的修正之后,下一步是利用先进的数据科学方法,理解并降低随机误差(random error)对水溶解度模型拟合的影响。目前文献中常用的拟合方法是使用基于最小二乘法的简单线性回归对现有的数据集进行分析(Mierdel et al., 2007; Litasov et al., 2011),这些拟合结果往往不能处理像水逸度这样较强的非线性关系,同时少数异常点(outlier)的存在会严重影响拟合结果的稳定性。Dong 等(2021)使用基于迭代重加权最小二乘法(iterated reweighted least squares)的稳健非线性回归(nonlinear robust regression)来同时拟合水逸度、温度、压强和铁含量对 NAMs 水溶解度的影响。迭代重加权最小二乘法是一种替代最小二乘法的稳健拟合方法,相比之下,它可以帮助我们识别并且减轻少数异常点对拟合结果的影响(Maronna et al., 2019)。

通过使用修正后的实验数据集对前文推导的热力学模型进行拟合,笔者对地球地幔的储水能力进行了计算。地球地幔矿物的成分主要为各矿物的镁端元相,它们的 $Mg^\#$ 主要在85~95之间。基于富镁($Mg^\# = 81~100$)的橄榄石、瓦兹利石和林伍德石的数据集,Dong 等(2021)拟合公式(8)中的五个参数 $a$、$b$、$c$、$d$ 和 $n$。对于橄榄石来说,$a$、$c$ 和 $d$ 被发现没有显著的统计学意义[$p>0.01$,方程(12)],说明在 $Mg^\# = 81~100$ 范围内,$X_{Fe}$ 对其水溶解度没有明显影响,而 $P$ 通过改变 $f_{H_2O}$ 项影响其水溶解度(图1a;表1):

$$\ln(c_{H_2O}) = \frac{n}{2} \cdot \ln(f_{H_2O}) + \frac{b}{T} \tag{12}$$

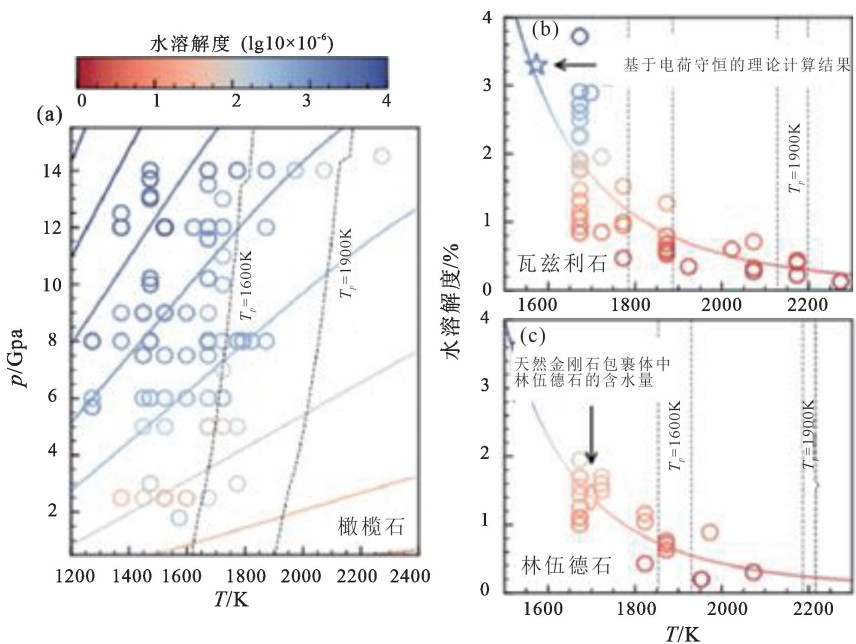

实验数据和拟合模型分别由彩色圆圈和彩色线条表示。在(a-c)中,黑色虚线和矩形框表示现代和太古宙地幔温度
($T_p$ = 1600 K, $T_p$ = 1900 K)。本图修改自 Dong 等(2021)

图 1 (a)橄榄石,(b)瓦兹利石,和(c)林伍德石在高温高压下的水溶解度

Fig. 1 Water storage capacities in (a) olivine, (b) wadsleyite, and (c) ringwoodite at high pressure and temperature

与此同时,Dong 等(2021)发现,对瓦兹利石和林伍德石而言,$n$、$c$ 和 $d$ 在统计学上不显著($p > 0.01$,即 $f_{H_2O}$、$P$ 和 $X_{Fe}$ 对水溶解度没有显著影响)。通过稳健回归对 $a$ 和 $b$ 的拟合(图 1b 和 1c),瓦兹利石和林伍德石的水溶解度模型可以简化为方程(13):

$$\ln(c_{H_2O}) = a + \frac{b}{T} \tag{13}$$

读者可以在表 1 中查找适用于地球地幔橄榄石、瓦兹利石和林伍德石的水溶解度模型参数(图 1,表 1)。

表 1 橄榄石、瓦兹利石和林伍德石的水溶解度的稳健拟合模型

Table 1 Coefficients of the water storage capacity parameterizations for olivine, wadsleyite, and ringwoodite

| 矿物名称 | 具有统计学意义的热力学参数(标准差)* | |
|---|---|---|
| 橄榄石 | n | b |
|  | 0.6447 (0.0483) | 4905.5403 (416.2888) |
| 瓦兹利石 | a | b |
|  | -7.6356 (0.8952) | 13739.5371 (1602.2032) |
| 林伍德石 | a | b |
|  | -6.8856 (1.3651) | 12206.2676 (2400.7947) |

在地幔中,布里奇曼石占其质量的一半以上。因此,计算地球地幔整体储水能力的关键是布里奇曼石的水溶解度。遗憾的是,目前关于布里奇曼石水溶解度的实验数据十分有限。其原因主要是,在高压实验中,在达到下地幔压强条件下,控制成分和起始材料的含水量并同时获得足够大的单晶来进行含水量测量是非常困难的。在现有的实验数据集中,即使在相似温压下用相同方法测得的布里奇曼石水溶解度误差也仍然高达 3 个数量级(Meade et al., 1994;Bolfan-Casanova et al., 2000;Murakami et al., 2002;Litasov et al., 2003;Inoue et al., 2010;Fu et al., 2019;Liu et al., 2021;图 2)。异常高的值可能是由微米或纳米级的含水矿物包裹体造成的,而异常低的值则可能是由于起始材料中的水不足而导致的。另外,这些实验数据都是在下地幔顶部的较低的压强范围内(20~30 GPa)测得的,因此,不仅现有的少量数据不足以拟合像公式(8)这样复杂的热力学模型,更无法给布里奇曼石水溶解度和压强的关系提供有效的实验约束。从橄榄石、瓦兹利石和林伍德石的拟合结果中可以看出,温度是另一个可能对布里奇曼石水溶解度有较大影响的变量。于是,在假设布里奇曼石水溶解度不受压强变化影响的前提下,Dong 等(2021)使用林伍德石和布里奇曼石之间的水分配系数($D_{rw/bdg}$)估算了布里奇曼石在不同温度下的水溶解度,然后运用蒙特卡洛方法来估计其对地球地幔储水能力不确定性的影响。在

使用由实验确定的 $D_{rw/bdg}$ 的值和其误差 [$D_{rw/bdg}$ = 15±8; Inoue 等(2010)]来进行随机蒙特卡洛取样后,布里奇曼石水溶解度是 $370^{+1240}_{-180} \times 10^{-6}$(25 GPa 和 1953 K,沿 1600 K 绝热线)。如果把用蒙特卡罗方法估计的布里奇曼石水溶解度与文献实验数据比较,现有的实验结果几乎都落在模型的不确定性区间内(5~95 百分位距:$190 \times 10^{-6}$ ~ $1610 \times 10^{-6}$,25 GPa,2230 K,沿 1900 K 绝热线)(图 2)。如果进一步假设下地幔的剩余的两个矿物——毛钙硅石和铁方镁石,分别可以储存 $10 \times 10^{-6}$ 的水,那么地球地幔的储水能力是 $290^{+950}_{-140} \times 10^{-6}$(25 GPa,1953 K,沿 1600 K 绝热线)。本节中各矿物水溶解度和地幔储水能力的误差均取模拟结果的 5 至 95 百分位距(图 2)。

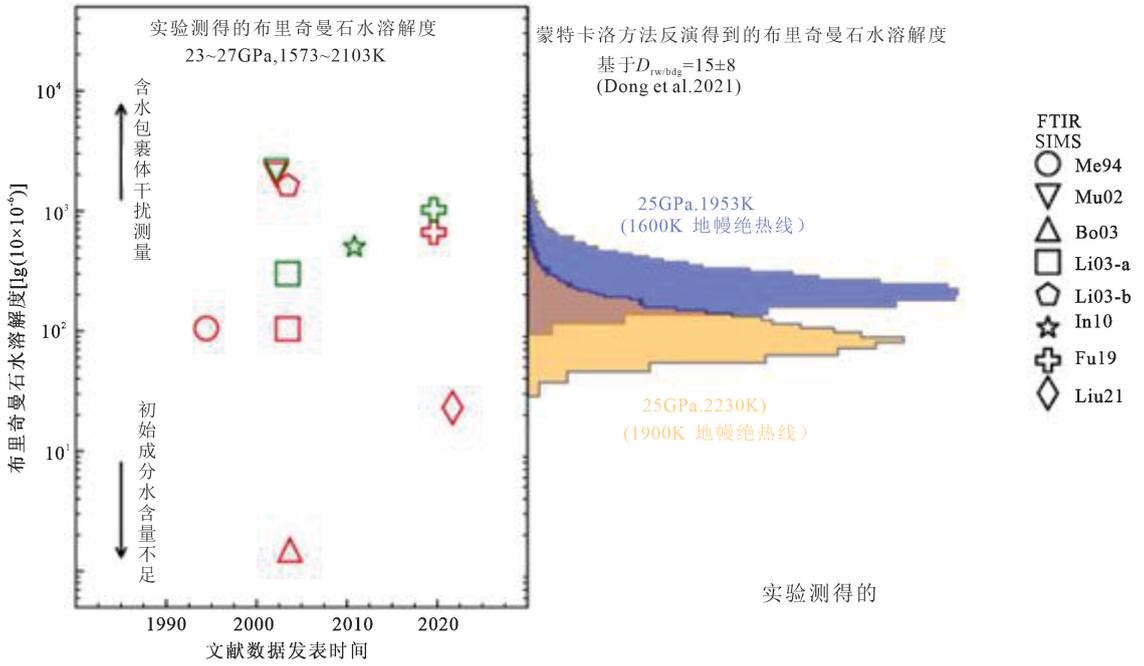

现有文献中数据误差高达 3 个数量级:从 ~$10 \times 10^{-6}$ 到 $1000 \times 10^{-6}$(Meade et al.,1994;Bolfan-Casanova et al.,2000;Murakami et al.,2002;Litasov et al.,2003;Inoue et al.,2010;Fu et al.,2019;Liu et al.,2021)。蓝色(25 GPa 和 1953 K,沿 1600 K 绝热线)和黄色(25 GPa,2230 K,沿 1900 K 绝热线)的柱状图是在使用由实验确定的 $D_{rw/bdg}$ 的值和其误差来来进行随机蒙特卡洛模拟的结果

图 2 实验测得的布里奇曼石水溶解度的汇编

Fig. 2 Compilation of experimentally measured water solubility in bridgmanite

## 3 基于自助聚集算法的火星地幔储水能力的模型

与地球相比,火星地幔含有较多的铁,其 Mg# 约为 75(Taylor,2013),而地球地幔的 Mg# 则接近 89(Workmana and Hart,2005)。在围绕地球地幔的研究中,Dong 等(2021)发现铁对橄榄石、瓦兹利石和林伍德石水溶解度的影响在 Mg# = 81~100($X_{Fe}$ = 0~0.19)的成分区间内没有统计显著性,即方程(8)中的成分项 $\frac{d \cdot X_{Fe}}{T}$ 的参数 $d$ 的 $p$ 值大于 0.01。值得注意的是,$p$ 值的显著性区间($p$ = 0.01)仅仅是一个约定俗成的经验标准(Wasserstein and Lazar,2016),在 $X_{Fe}$ < 0.19 的范围内,关于 $d$ 的 $p$ 值大于 0.01 并不代表铁含量对这些矿物的水溶解度没有影响。该发现只能说明,现有的 $X_{Fe}$ < 0.19 的数据不足以判定铁含量的影响。而富铁的火星矿物 $X_{Fe}$ 范围约为 0.2~0.4,远高于地球地幔矿物的 $X_{Fe}$,因此,在火星地幔中铁对矿物水溶解度的影响则因另当别论。

从大压机高压实验结果里看出,铁在 $X_{Fe}$ > 0.2 时对橄榄石水溶解度有明显的影响。当橄榄石的 $X_{Fe}$ 从 0.2 增加到 0.4 时(3~6 GPa),其水溶解度会增加约 2~3 倍(Withers et al.,2011)。对瓦兹利石和林伍德石而言,目前还缺乏铁含量对它们水溶解度的影响的系统性研究。虽然 Dong 等(2021)没有发现铁含量对它们水溶解度有显著的统计学意义($p$>0.01),但是把根据地球矿物($X_{Fe}$ = 0~0.19)得出的结论推广到富铁的火星矿物中去($X_{Fe}$ = 0.2~0.4),则可能会引起较大误差。于是,Dong 等(2022)在 Dong 等(2021)汇编的数据集中,加入了

相关的富铁 NAMs 实验数据（$X_{Fe}>0.19$）。因为对富铁橄榄石、瓦兹利石和林伍德石的实验约束十分有限，所以，即使在使用扩充后的数据集来稳健拟合方程(8)中的 $\dfrac{d \cdot X_{Fe}}{T}$ 项，$d$ 的 $p$ 值仍大于 0.01，因此可能会造成了对 $d$ 的过度拟合(图3)。

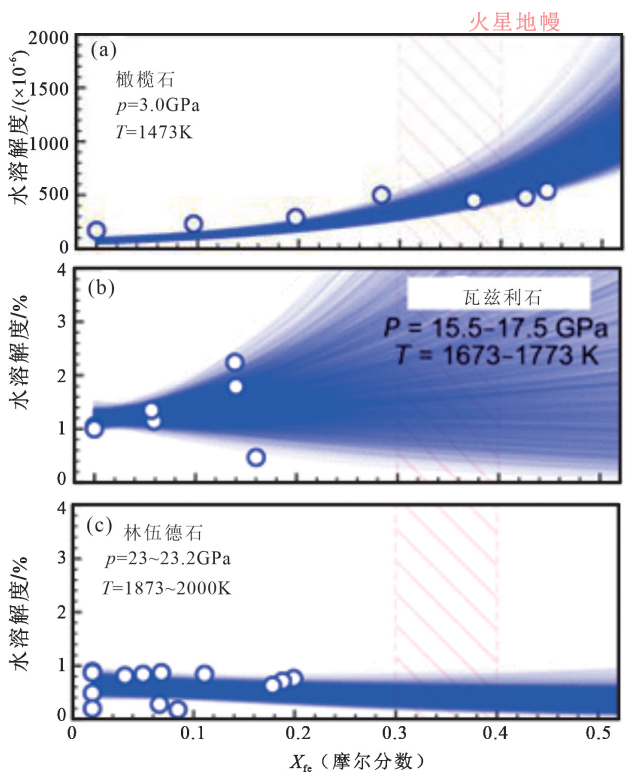

实验数据由圆圈表示。曲线对应的是，在特定温压下，$10^4$ 个稳健回归拟合模型。
红色阴影区域表示火星地幔矿物铁含量的大致范围。本图修改自 Dong 等(2022)

图 3　铁含量对(a)橄榄石、(b)瓦兹利石和(c)林伍德石的水溶解度的影响

Fig. 3　Effects of Fe content on the water solubilities of (a) olivine, (b) wadsleyite, and (c) ringwoodite at selected pressures and temperatures

为了避免过度拟合 $d$，同时也将扩充后所有实验数据集纳入我们对火星地幔储水能力及其不确定性的估算中，Dong 等(2022)应用了一种叫做自助聚集(bootstrap aggregation)或"装袋"(bagging)的集成学习算法(ensemble learning)。首先对三种火星地幔主要矿物(橄榄石、瓦兹利石和林伍德石)的水溶解度的实验数据集进行了重复随机替换抽样，从而生成 $3\times10000$ 个新的样本数据集。然后，用方程 8 对这些样本数据集进行稳健非线性拟合(参看本文第二节)，从而得到 $3\times10000$ 个水溶解度拟合模型(图 3)。由于重抽样后的样本集各不相同，其拟合出的温度、压强、铁含量与水溶解度的关系也会不同。比如说，对于橄榄石而言，因为其富铁的实验数据数量相对充足且自洽性好，所以拟合出的 10000 个水溶解度模型都显示其铁含量与水溶解度正相关(图 3a)。相反，对于瓦兹利石而言，因为其富铁的实验数据数量有限且自洽性差，所以，拟合出的铁含量影响有时与水溶解度正相关，有时负相关(图 3b)。接着，计算地幔各个矿物的水溶解度的加权和，生成 10000 个火星地幔储水能力模型。最后，通过取所有模型的平均值获得最优地幔储水能力模型。该"装袋"方法因为平均了 10000 个火星地幔储水能力模型，$M_1, M_2, \cdots, M_n$(各模型的方差为 $\sigma^2$)，所以，最优地幔储水能力模型 $\overline{M}$ 的方差仅为 $\sigma^2/N$，从而避免过拟合的发生(James et al., 2021)。这种"装袋"方法与 Dong 等(2021)使用的蒙特卡洛方法不同，它不但不需要在拟合模型中明确铁含量的影响，而且可以将 $\dfrac{d \cdot X_{Fe}}{T}$ 的参数 $d$ 的不确定性直接纳入对火星地幔储水能力的不确定性计算中。尽管方法不同，通过自助聚集算法发现的压强和温度对水溶解度的影响，与由蒙特卡洛方法得出的结果仍然一致(图4)。

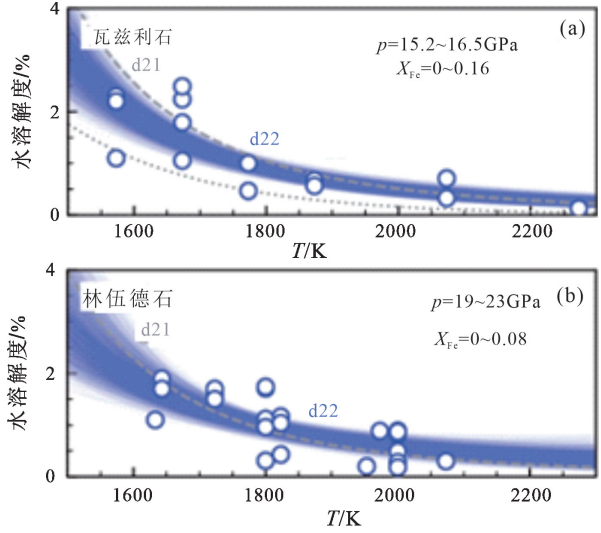

实验数据由圆圈表示,蓝色曲线对应对 $10^4$ 个重抽样后样本集的稳健非线性拟合(Dong et al., 2022);虚线对应 Dong 等(2021)中对富镁数据($Mg^\# = 81 \sim 100$)的单一稳健拟合结果;点虚线对应 Litasov 等(2011)中对镁端元数据($Mg^\# = 100$)的单一指数函数拟合。本图修改自 Dong 等(2022)

图 4 温度对(a)瓦兹利石和(b)林伍德石水溶解度的影响

Fig. 4 Effects of temperature on the water solubilities of (a) olivine, (b) wadsleyite, and (c) ringwoodite at selected pressures and temperatures

## 4 火星和地球固体地幔的储水能力和深部水循环

根据温度、压强和矿物丰度等,(Dong et al., 2021,2022)先沿着不同的绝热地温梯度(地幔潜在温度 $T_p = 1500 \sim 1900$ K)计算出地球和火星地幔储水能力曲线(图 5a、5b),再沿着地温梯度对地幔储水能力进行积分,得到在特定 $T_p$ 下地幔整体储水能力(图 5c)。不论是地球还是火星,它们的上地幔的储水能力都会随着压强增加而增加。对于地球而言,在上地幔和过渡带的交界处,他们的储水能力相差 4~6 个数量级(~13~15 GPa),从而在 410 公里不连续处形成了地幔储水能力的跃变(图 5b)。相比之下,因为铁会增加橄榄石水溶解度,所以,在相似温压下,火星上地幔($Mg^\# \approx 75$)可以储存比地球上地幔($Mg^\# \approx 89$)更多的水,而且火星地幔上地幔和过渡带的交界处也没有储水能力的跃变。

如果以地球表面的海洋作为单位(一个地球海洋的质量,1 OM = $1.335 \times 10^{21}$),现代地球地幔的整体储水能力为 $2.29^{+2.12}_{-0.43}$ OM(沿 1600 K 绝热线),太古宙时期的地球地幔的整体储水能力为 $0.72^{+0.97}_{-0.20}$ OM(沿 1900 K 绝热线,太古宙 $T_p$ 比现代 $T_p$ 高 300 K)。对于火星,人们通常使用全球等效层(GEL)来衡量其水含量,该单位代表一定量的水如果均匀地分布在整个火星上所能形成的水圈的厚度。现代火星地幔的整体储水能力为 $9.0^{+2.8}_{-2.2}$ km GEL(沿 1600 K 绝热线)。如果火星地幔 $T_p$ 在诺亚纪时期(约 4.1Ga;Filiberto,2017))比其现代地幔 $T_p$ 高 200 K,那么火星地幔整体储水能力将会减少为 $6.1^{+1.9}_{-1.7}$ km GEL。本节中地幔整体储水能力的误差取模拟结果的 5 至 95 百分位距。

地球和火星地幔整体储水能力都随着 $T_p$ 的降低而增加(图 5c),这主要是由于温度与三种地幔主要矿物(橄榄石、瓦兹利石、和林伍德石)的水溶解度都成负相关。有意思的是,尽管地球地幔的体积(~$9.1 \times 10^{11}$ km³)是火星地幔(~$1.4 \times 10^{11}$ km³)的 ~6.5 倍,其整体储水能力(~2.3 OM)却只是火星地幔(~1.0 OM)的 2.3 倍,这一比例失调主要归咎于火星地幔中的富铁橄榄石的高水溶解度。另外,地球和火星地幔的整体储水能力在 $T_p$ 较高时趋于一致的,其原因有二:一是温度对 NAMs 水溶解度的影响超过了铁含量对其的影响;二是当 $T_p$ 较高时,地球下地幔储水能力在对整体地幔储水的贡献被明显削弱了。

需要注意的是,虽然地球和火星的地幔整体储水能力都在演化冷却的过程中增加,但是两个行星深部的水循环历史可能大相径庭。在地球上,自板块运动开始后,俯冲作用就会将地表水重新带回地幔深处,不断增加其地幔实际含水量(Van Keken et al.,2011;Dong et al.,2021;图 5c 中浅红色虚线

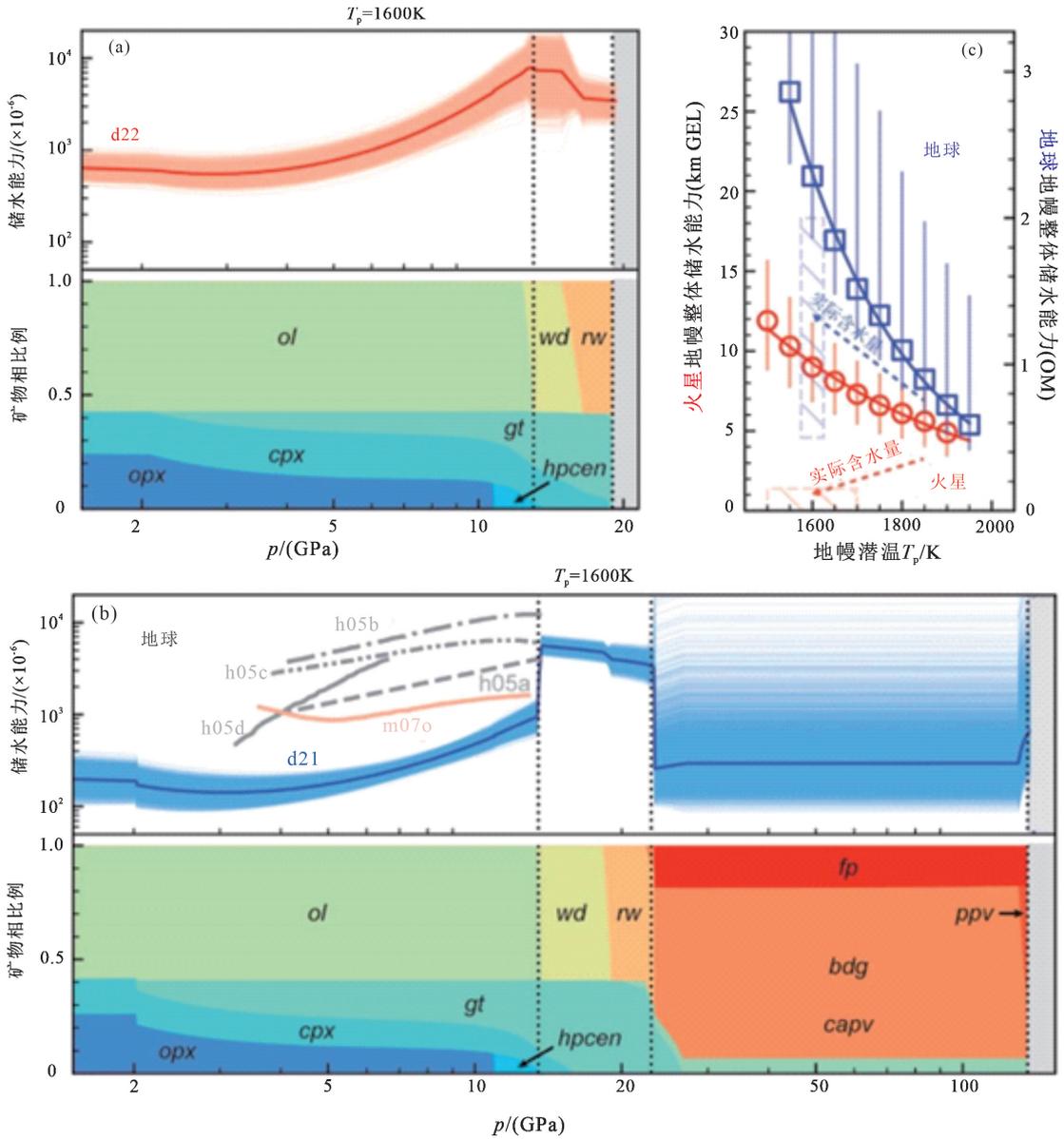

在(b)中，灰色曲线是 Hirschmann 等(2005)的储水能力曲线；红色曲线是 Mierdel 等(2007)的储水能力曲线；在(c)中，红色和蓝色圆圈分别是火星地幔和地球地幔在各个 $T_p$ 下地幔整体储水能力；阴影区域代表火星(红色，14 ppm wt～250 ppm wt×$10^{-6}$ 或 80～1460 m GEL；Filiberto et al.，2018)和地球(蓝色，0.5～2 OM 或 5～19km GEL；Dauphas 和 Morbidelli，2014；Hirschmann，2018)地幔实际含水量的地球化学约束

图 5 当今火星地幔(a)和地球地幔(b)的储水能力曲线和平衡矿物相组合($T_p$ = 1600 K)，以及它们分别的整体储水能力与地幔潜温($T_p$)的关系。

Fig. 5 Water storage capacity profiles and equilibrium mineral phase assemblies ($T_p$ = 1600K) of (a) present-day Mars mantle and (b) present-day Earth mantle, and (c) their respective overall water storage capacity versus mantle potential temperature ($T_p$)

箭头)。相反，火星因为缺乏像俯冲作用这样有效的深部水循环机制，其地幔在早期火山作用排气时就失去了大部分水，并且实际含水量一直保持在相对较低的水平(Dong et al.，2022；图 5c 中浅红色虚线箭头)。

火星和地球地幔的实际含水量可以从天然样品进行推断，如火星陨石、洋岛玄武岩、地幔捕虏体和金刚石包裹体：根据火星陨石样本可以推断出其地幔含水量为 0.8～1 km GEL 或 200×$10^{-6}$～300×$10^{-6}$(Filiberto et al.，2018)；根据洋岛玄武岩样本可以推断地球地幔含水量不超过 2 OM 或 700×$10^{-6}$～800×$10^{-6}$(Peslier et al.，2017)。然而，这些估计值都假设它们地幔水含量分布均一，实际上，大多数这些天然样品只来自浅层地幔，并不一定能代表深

部地幔的含水量。同时,即使在同一深度,地幔水的分布也可能有较大不均一性,比如说,含有俯冲板块的地球地幔过渡带往往更富水(Pearson et al., 2014)。因此,对火星和地球地幔实际水含量的估计仍有很大争议。对地球地幔而言,还可以通过电导率(Kelbert et al., 2009)、粘度(Fei et al., 2017)和剪切波速度(Schulze et al., 2018;Wang et al., 2019,2021)等地球物理方法推断其特定区域的实际含水量。基于本文的主旨是讨论地幔储水能力,笔者略去对地幔实际含水量研究的全面回顾,感兴趣的读者可以参照 Filiberto 等(2018)和 Peslier 等(2017)。(图5)

## 5 岩石系外行星地幔储水能力的统计性质

随着系外行星观测研究的蓬勃发展,对岩石行星的研究不再拘泥于对地球和火星这两个太阳系内岩石行星的"案例研究"。根据 Dong 等(2021)对地幔主要矿物水溶解度的稳健拟合结果,Guimond 等(2023)将其为地球建立的地幔储水能力模型推广成为一个对系外行星广泛适用的通用模型,并进一步地分析了 Hypatia 数据集中恒星周围数以百计的系外行星的统计性质(Hinkel et al., 2014)。其中,行星地幔 Mg/Si,行星地幔温度和行星质量是决定一颗系外行星地幔储水能力的关键参数(图6)。在对 FGKM 恒星周围的岩石行星进行统计分析后,Guimond 等(2023)发现,在0.1到3个地球质量之间,行星地幔 Mg/Si 和地幔温度对其地幔储水能力造成的变化大约是 0.3~13 OM(图7)。同时,在 0.3~3 个地球质量之间($M_\oplus$),系外行星的质量($M_p$)和其地幔储水能力($C_p$)的经验关系可以用公式14表达:

$$C_p = M_p^{0.69} \quad (14)$$

需要强调的是,这些系外行星地幔储水能力的统计性质往往有其矿物学或者地球物理相关的解释。岩石行星地幔中 Mg/Si 制约着其矿物相组成。

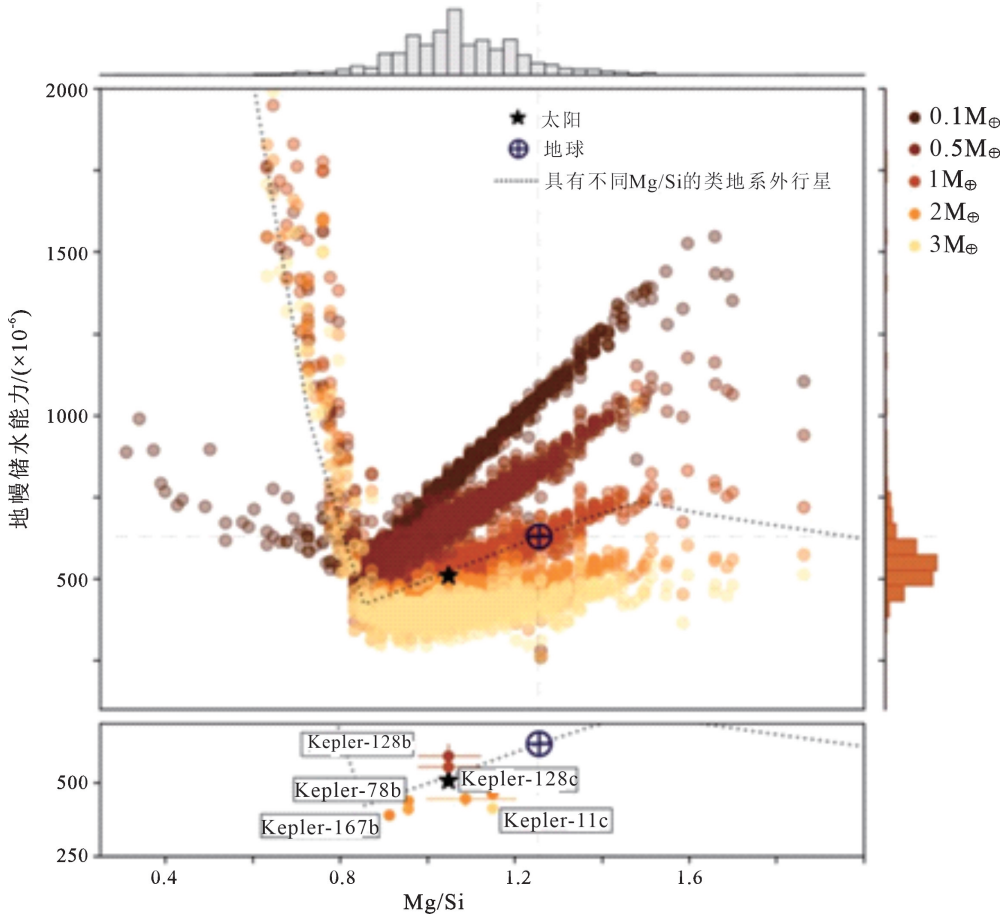

(a)在 Hypatia 目录中宿主恒星周围的系外行星;(b)开普勒空间望远镜观测确认的系外岩石行星的地幔成分。
本图修改自 Guimond 等(2023)

图6 岩石行星地幔储水能力和地幔成分的关系

Fig. 6 Relationship between water storage capacity of the rocky planetary mantle and its composition

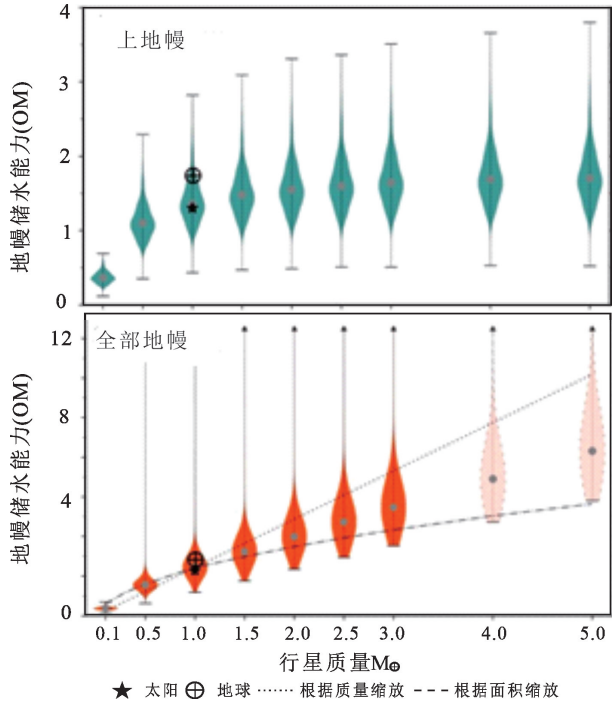

误差线代表在相同行星质量时,其地幔储水能力随地幔成分的变化。⊕代表地球成分,
实心圆圈代表宿主恒星成分。本图修改自 Guimond 等(2023)

图 7 岩石行星地幔储水能力和行星质量的关系（a：上地幔；b：地幔整体）

Fig. 7 Relationship between water storage capacity of the rocky planetary mantle and planetary mass (a: upper mantle; b: whole mantle)

比如说,提高 Mg/Si 往往会增加橄榄石和石榴子石的比例,或者降低林伍德石的比例(Dong et al., 2022；Guimond et al., 2023)。随着地幔矿物相比例的变化,系外岩石行星地幔的储水能力也会随之改变。在缺乏对其地幔成分直接观测的情况下,Guimond 等(2023)通过宿主恒星的 Mg/Si 来估算其星系内岩石行星的 Mg/Si。在 Hypatia 数据集中,系外行星地幔的 Mg/Si 主要落在 0.72～1.41 的区间内。这个范围看似不大,但是其对地幔矿物相比例的影响十分显著:当 Mg/Si 接近 1 时,橄榄石和辉石的比例也接近 1。但当 Mg/Si 接近 0.7 时,过渡带中大部分的瓦兹利石和林伍德石基本都被石榴子石取代。根据 Hypatia 数据集中的 Mg/Si 可能的变化,Guimond 等(2023)发现其模型估算出的上地幔储水能力会相差两倍之多。值得注意的是,虽然 Guimond 等(2023)的模型并没有将铁的影响包括在其通用模型中,但是在相同的 Mg/Si 条件下,富 Fe 的岩石行星地幔[以 Dong 等(2022)对火星的研究为例]会使其橄榄石的水溶解度增高,因此,其储水能力可能比贫 Fe 的地幔更高。另一个制约系外行星地幔矿物相组成的关键因素是行星质量。在大于一个地球质量的系外岩石行星地幔中,主要的矿物可能不再是上地幔的橄榄石或者是下地幔的布里奇曼石。当地幔压强超过 130～150 GPa 时,地幔将主要由后钙钛矿构成,如果压强继续增高,后钙钛矿可能会进一步分解为 MgO 和 $SiO_2$ 的超高压同质异形体(Duffy et al., 2015)。后钙钛矿和 $SiO_2$ 在超高压下,水溶解度可以达到几千×$10^{-6}$,甚至更高(Townsend et al., 2016；Lin et al., 2022)。包括后钙钛矿在内这些超高压矿物相将取代布里奇曼石,成为超级地球(大于三个地球质量的系外岩石行星)地幔中的主要矿物。但由于当前缺乏对这些超高压矿物相水溶解度的实验约束,Guimond 等(2023)的模型将不再适用(图 7)。

同时,地幔温度对岩石行星地幔的储水能力也有较大影响。温度是在微观层面上控制 NAMs 水溶解度的重要参数。升高温度通常会降低几乎所有 NAMs 水溶解度。在行星质量和成分类似时,温度较高(年轻的)的岩石行星地幔的储水能力会比温度较低(古老的)的岩石行星地幔低得多。Guimond 等(2023)使用的 Hypatia 数据集并没有对行星年龄或者内部温度的直接约束,因此,其通用模型的误差,有一部分源于我们对这些行星内部温度的不了解。

当我们把系外行星铁核中以铁氢合金形式存在的水也计算在内时,其行星整体的储水能力可能高达其质量的 6%,相当于 800 公里的表面海洋(Shah et al., 2021)。在 $0.1 \leq M/M_\oplus \leq 3$ 的质量范围内,当一个岩石行星将其内部储存的水全部释放到其表面时,其半径的变化可以达到 5%。在现有的和未来精度更高的系外行星观测任务中,如 JW-ST、PLATO 等(Hatzes, 2014),我们或许可以用未来观测到的系外行星的半径差异去估算他们内部的实际含水量。

# 6 总结与展望

本文主要探讨了数据科学方法在估计岩石行星地幔储水能力研究中的应用。在已知布里奇曼石水溶解度等参数的误差分布时,第三节中介绍的蒙特卡洛方法可以对地球地幔储水能力模型的不确定性进行定量分析。同一节中介绍的稳健回归方法,可以解决传统回归方法中对异常值敏感、容易过拟合等问题。尤其是在处理像水溶解度测量这类随机误差较大的数据集,稳健回归方法可以帮助我们识别和减弱异常值对最终拟合结果的影响。第四节中的研究铁含量和 NAMs 水溶解度关系的案例告诉我们,当实验数据过于稀疏以至于无法确定具体热力学拟合模型时,如果我们运用像"装袋"这类统计学习算法,可在没有具体拟合模型时,把水溶解度实验数据点的测量误差直接转化成对火星地幔整体储水能力不确定性的估算中去。

数据科学的核心是数据。像水溶解度这类矿物物理学数据集,其主要数据来源是高压实验。这些数据集相对较小,数据之间误差大、自洽性低。在这种的情况下,即使利用较为先进的统计和统计学习方法,还是存在对数据过拟合的风险。因此,如果想要进一步理解地幔储水能力,缩小目前模型不确定性,我们还是需要从实验数据本身入手。比如准确测量布里奇曼石、$SiO_2$ 后钙钛矿等超高压矿物在不同温压下的水溶解度,或是系统研究铁含量对 NAMs 水溶解度的影响。对现有矿物物理数据的统计研究可以看作是一个对数据挖掘的过程,它可以帮助我们有效识别现有数据集中哪些温度、压强或者成分的数据误差大、自洽性低,从而作为实验者的参考,成为设计未来实验的路线图。当然,获取误差小、自洽性高的矿物物理学数据集并非一朝一夕可以达成,最终还是取决于高压实验技术的进步。与此同时,快速增长的系外岩石行星样本为我们用数据科学研究矿物物理问题提供了另一个研究对象。当矿物物理数据数量和质量有限时,即使不能定量地约束地球和火星的复杂模型,我们仍可以用较为定性的模型去研究系外行星的统计性质,从而帮助我们更为全面地理解岩石行星地幔储水能力。

目前,运用大数据和数据科学方法去研究矿物物理相关的问题的尝试还比较少。因此,讨论这三个例子只是抛砖引玉,笔者希望,更多学者可以将大数据和数据科学方法运用到矿物物理学中,给我们对行星内部结构、成分和演化的研究提供新思路。